\documentclass[aps,pra,showpacs,superscriptaddress,twocolumn,reprint]{revtex4-1}

\usepackage{amsmath,amsfonts,amssymb}
\usepackage{graphicx}
\usepackage{epstopdf}
\usepackage{bm}

\begin{document}
\title{Detecting quantum backflow by the density of a Bose-Einstein condensate}

\author{M. Palmero}
\affiliation{Departamento de Qu\'imica F\'isica, Universidad del Pa\'is Vasco UPV/EHU, Apdo. 644, Bilbao, Spain}

\author{E. Torrontegui}
\affiliation{Departamento de Qu\'imica F\'isica, Universidad del Pa\'is Vasco UPV/EHU, Apdo. 644, Bilbao, Spain}

\author{J. G. Muga}
\affiliation{Departamento de Qu\'imica F\'isica, Universidad del Pa\'is Vasco UPV/EHU, Apdo. 644, Bilbao, Spain}
\affiliation{Department of Physics, Shanghai University, 200444 Shanghai, People's Republic of China}

\author{M. Modugno}
\affiliation{{Dpto.~de F\'isica Te\'orica e Hist.~de la Ciencia, Universidad del Pa\'is Vasco UPV/EHU, 48080 Bilbao, Spain}}
\affiliation{IKERBASQUE, Basque Foundation for Science, Alameda Urquijo 36, 48011 Bilbao, Spain}
\begin{abstract}
Quantum backflow is a classically forbidden effect consisting in a negative flux for states with negligible negative momentum components. It has never been  observed experimentally so far. We derive a general relation that connects backflow with a critical value of the particle density, paving the way  for the detection of backflow by a density measurement. To this end,  we propose an explicit scheme with Bose-Einstein condensates, at reach with current experimental technologies. Remarkably, the application of a positive momentum kick, via a Bragg pulse, to a condensate with a positive velocity may cause a current flow in the negative direction.  
\end{abstract}
\maketitle

\section{introduction}

Quantum backflow is a fascinating quantum interference effect 
consisting in a negative current density for quantum wave packets without negative momentum components \cite{allcock}. It reflects a fundamental point about quantum measurements of velocity: usual measurements of momentum (velocity) distributions are performed globally - with no resolution in position - whereas the detection of the velocity field (or of the flux) implies a local measurement, that may  provide values outside the domain of the global 
velocities, due to non commutativity of momentum and position \cite{local}.
Despite its intriguing nature - obviously counterintuitive from a classical viewpoint - quantum backflow has not yet received as much attention as other quantum effects.
Firstly discovered by Allcock in 1969 \cite{allcock}, it only started to be studied in the mid  90's.  Bracken and Melloy \cite{BM} provided a bound for the maximal fraction of probability that can undergo backflow. 
Then, additional bounds and analytic examples, and its implications in the definition of arrival times of quantum particles where discussed by Muga \textit{et al.} \cite{muga,Leav,Jus}. Recently, Berry \cite{BerryBack} analyzed the statistics of backflow for random wavefunctions, and  Yearsley \textit{et al.} \cite{year} studied some specific cases, clarifying the maximal backflow limit. 
However, so far no experiments have been performed, and a clear program to carry out  one is also missing. 
Two important challenges  are the measurement of the current density (the existing proposals  for local and direct 
measurements are rather idealized schemes \cite{Jus}), and the preparation of states with a detectable amount of backflow.  

In this paper we derive a general relation that connects the current and the particle density, allowing for the detection of backflow by a density measurement, and 
propose a scheme for its observation with Bose-Einstein condensates in harmonic traps,
that could be easily implemented with current experimental technologies.   
In particular, we show that preparing a condensate with positive-momentum components, and then further transferring a positive momentum kick to part of the atoms, causes under certain conditions, remarkably, a current flow in the negative direction. 
Bose-Einstein condensates are particularly promising for this aim because, besides their high level of control and manipulation, they are quantum matter waves where the probability density and flux are in fact a density and flux of particles --in contrast to a statistical ensemble of single particles 
sent one by one--, and in principle allow for the measurements of local properties in a single shot experiment.

\begin{figure}[t]
\centerline{\includegraphics[width=0.9\columnwidth]{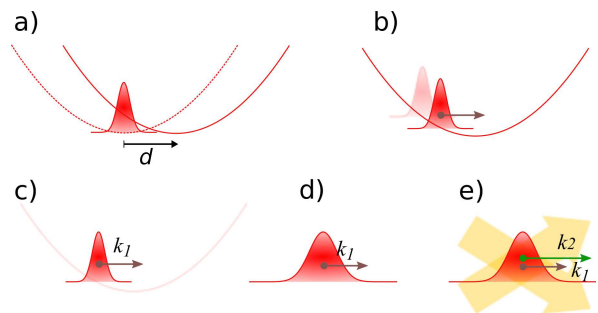}}
\caption{(Color online) a) A condensate is created in the ground state of a harmonic trap with frequency $\omega_{x}$; at $t=0$ we apply a magnetic gradient that shifts the trap by a distance $d$; b) the condensate starts to perform dipole oscillations in the trap; c) when it reaches a desired momentum 
$\hbar k_{1}$ the trap is switched off, and d) then the condensate is let expand for a time $t$; e) finally, a Bragg pulse is applied in order to transfer part of the atoms to a state of momentum $\hbar k_{2}$.}
\label{fig:scheme}
\end{figure}

\section{general scheme}
Let us start by considering a one dimensional Bose-Einstein condensate with a narrow momentum distribution centered around $\hbar k_{1}>0$, with negligible negative components. 
Then, we apply a Bragg pulse that transfers a momentum $\hbar q>0$ to part of the atoms \cite{bragg}, populating a state of momentum $\hbar k_{2}=\hbar k_{1} + \hbar q$ (see Fig. \ref{fig:scheme}). By indicating with $A_{1}$ and $A_{2}$ the amplitudes of the two momentum states, 
the total wave function is
\begin{equation}
\Psi(x,t)=\psi(x,t)\left(A_{1} + A_{2}\exp\left[iq x + i\varphi\right]\right),
\label{eq:bragg}
\end{equation}
where we can assume $A_{1}, A_{2}\in \mathbb{R}^{+}$ without loss of generality (with $A_{1}^{2}+A_{2}^{2}=1$), $\varphi$ being an arbitrary phase. All these parameters (except the phase, that will be irrelevant in our scheme), 
can be controlled and measured in the experiment. Then, by writing the wave function of the initial wave packet  as
\begin{equation}
\psi(x,t)=\phi(x,t)\exp[i\theta(x,t)]
\end{equation}
with $\phi$ and $\theta$ being real valued functions,
the expression for the total current density, $J_{\Psi}(x,t)
=(\hbar/m)\textrm{Im}\left[\Psi^{*}\nabla\Psi\right]$ can be easily put in the form
\begin{equation}
\frac{m}{\hbar}J_{\Psi}(x,t)
=(\nabla\theta)\rho_{\Psi}+\frac12q \left[\rho_{\Psi}+|\phi|^{2} (A_{2}^{2} -A_{1}^{2})\right],
\label{eq:current}
\end{equation}
with $\rho_{\Psi}(x,t)=|\phi(x,t)|^{2}\left(A_{1}^{2} + A_{2}^{2}+ 2A_{1}A_{2}\cos\left(qx + \varphi\right)\right)$ being the total density.
Therefore, a negative flux, $J_{\Psi}(x,t)<0$, corresponds to the following inequality for the density
\begin{equation}
\textrm{sign}[\eta(x,t)]\rho_{\Psi}(x,t)<
\frac{1}{|\eta(x,t)| }|\phi(x,t)|^{2} (A_{1}^{2} -A_{2}^{2}),
\end{equation}
where we have defined $\eta(x,t)= 1 +2\nabla\theta(x,t)/q$.
Later on we will show that  $\eta(x,t)<0$ corresponds to a \textit{classical regime}, whereas for $\eta(x,t)>0$ the backflow is a purely quantum effect, without any classical counterpart. 
Therefore, in the \textit{quantum regime}, backflow takes place when the density is below the following critical threshold: 
\begin{equation}
\rho_{\Psi}^{crit}(x,t)=
\frac{q}{q +2\nabla\theta(x,t) }|\phi(x,t)|^{2} (A_{1}^{2} -A_{2}^{2}).
\label{eq:denscrit}
\end{equation}
This is a fundamental relation that allows to detect backflow by a density measurement. It applies to any class of wavepackets of the form (\ref{eq:bragg}), including the superposition of two plane waves discussed in \cite{Jus,year}. We remark that, while in the ideal case of plane waves backflow repeats periodically at specific time intervals, for the present case 
it is limited to the transient when the two wave packets with momenta $\hbar k_{1}$ and $\hbar k_{2}$ are superimposed.

\section{An implementation with BEC\lowercase{s}}

\subsection{State preparation}

In order to propose a specific experimental implementation, we consider a condensate in a three dimensional harmonic trap, with axial frequency $\omega_{x}$. 
We assume a tight radial confinement, $\omega_{\perp}\gg\omega_{x}$, so that the 
wave function can be factorized in a radial and axial components (in the non interacting case this factorization is exact)\cite{note}. In the following we will focus on the
1D axial dynamics, taking place in the waveguide provided by the transverse confinement,  that is assumed to be always on. We define $a_{x}=\sqrt{\hbar/m\omega_{x}}$, $a_{\perp}=\sqrt{\hbar/m\omega_{\perp}}$. 

The scheme proceeds as highlighted in Fig.\ref{fig:scheme}.
The condensate is initially prepared in the ground state $\psi_{0}$ of the harmonic trap. Then, at $t=0$ the trap is suddenly shifted spatially by $d$ (Fig.\ref{fig:scheme}a) and the condensate starts to perform dipole oscillations (Fig.\ref{fig:scheme}b).
%
%
 Next, at $t=t_{1}$, when the condensate has reached a desired momentum $m v_{1}=\hbar k_{1}=\hbar k(t_{1})$, the trap is switched off (Fig.\ref{fig:scheme}c). 
%
%
At this point we let the condensate expand freely for a time $t$ (Fig.\ref{fig:scheme}d). 
Hereinafter we will consider explicitly two cases, namely a noninteracting condensate and the Thomas-Fermi (TF) limit \cite{dalfovo}, that can both be treated analytically. In fact, in both cases the expansion
can be expressed by a scaling transformation 
\begin{eqnarray}
\label{eq:scaling}
\psi(x,t)&=&\frac{1}{\sqrt{b(t)}}\psi_{0}\left(\frac{x-v_{1}t}{b(t)}\right)
\\
&\times&\exp\left[i\frac{m}{2\hbar }x^{2}
\frac{\dot{b}(t)}{b(t)} +i k_{1} x\left(1 -\frac{\dot{b}}{b}t\right) + i\beta(t)\right],
\nonumber
\end{eqnarray}
where $b(t)$ represents the scaling parameter 
and $\beta(t)$ is an irrelevant global phase (for convenience we have redefined time and spatial coordinates, so that at $t=0$ - when the trap is switched off - the condensate is centered at the origin). 
This expression can be easily obtained by generalizing the scaling in \cite{castin,kagan} to the case of an initial velocity field.

In the \textit{non interacting case}, the initial wave function is a minimum uncertainty Gaussian, 
$\psi_{0}(x)=({1}/{\pi^{\frac14}\sqrt{a_{x}}})\exp \left[-{x^2}/({2 a_{x}^2})\right]$,
and the scaling parameter evolves as $b(t)=\sqrt{1+\omega_{x}^{2}t^{2}}$ \cite{dalfovo,merzbacher}. 
For a TF distribution we have
$\psi_{0}(x)=\left[\left(\mu- \frac12 m\omega_{x}^{2}x^{2}\right)/g_{1D}\right]^{1/2}$
for $|x|<R_{TF}\equiv\sqrt{2\mu/m\omega_{x}^{2}}$ and vanishing elsewhere, with $g_{1D}=g_{3D}/(2\pi a_{\perp}^{2})$ \cite{salasnich}, and 
the chemical potential $\mu$ fixed by the normalization condition $\int\!dx|\psi|^{2}=N$, the latter being the number of atoms in the condensate \cite{dalfovo}. In this case $b(t)$ satisfies 
$\ddot{b}(t)=\omega_{x}^{2}/b^{2}(t)$,
whose asymptotic solution, for $t\gg1/\omega_{x}$, is 
$b(t)\simeq \sqrt{2} t \omega_{x}$ \cite{sp}.

Finally, we apply a Bragg pulse as discussed previously (Fig.\ref{fig:scheme}e). We may safely assume the duration of the pulse to be very short with respect to the other timescales of the problem \cite{bragg2}. 
Then, the resulting wave function is that in Eq. (\ref{eq:bragg}), with the corresponding critical density for backflow in Eq. (\ref{eq:denscrit}).   We have
\begin{equation}
\phi(x)=\frac{1}{\sqrt{b(t)}}\psi_{0}\left(\frac{x-(\hbar k_{1}/m)t}{b(t)}\right),
\end{equation}
while the expression for the phase gradient is
\begin{equation}
\nabla\theta=\frac{m}{\hbar }x
\frac{\dot{b}(t)}{b(t)} + k_{1}\left(1 -\frac{\dot{b}(t)}{b(t)}t\right)
\label{eq:grad-theta}
\end{equation}
that, in the asymptotic limit $t\gg1/\omega_{x}$, yields the same result $\nabla\theta={mx}/{\hbar t}$ for both the non interacting and TF wave packets.
Eventually, backflow can be probed by taking a snapshot of the interference pattern just after the Bragg pulse,  measuring precisely its minimum, and comparing it to the critical density. 

\subsection{Classical effects}

Before proceeding to the quantum backflow, let us discuss the occurrence of a classical backflow.
To this end it is sufficient to consider the flux of a single wave packet (before the Bragg pulse), namely $J_{\psi}(x,t) = (\hbar/m)|\phi|^{2}\nabla\theta$. In this case the flux is negative for $x<v_{1}(t-b/\dot{b})= :x_-(t)$ (see Eq. (\ref{eq:grad-theta})). Then, by indicating with $R_{0}$ the initial half width of the wave packet and with $R_{L}(t)=-b(t) R_{0} +v_{1}t$ its left border at time $t$, a negative flux occurs when $R_{L}<x_-$, that is for $R_{0}>v_{1}/\dot{b}(t)$. In the asymptotic limit, the latter relation reads $R_{0}>f v_{1}/\omega_{x}$ ($f=1,1/\sqrt{2}$ for the non interacting and TF cases, respectively). On the other hand, the momentum width of the wave packet is $\Delta_{p}\approx\hbar/R_{0}$ \cite{momentumwidth}, so that the negative momentum components can be safely neglected only when $mv_{1}\gg \hbar/R_{0}$. 
From these two conditions, we get that there is a negative flux (even in the absence of initial negative momenta) when $R_{0}\gg a_{x}$. This can be easily satisfied in the TF regime. In fact, in that case the backflow has a classical counterpart due to the force $F = -\partial_{x}(g_{1D}|\rho_{\psi}(x,t)|^{2})$ implied by the repulsive interparticle interactions \cite{castin}. 
These interactions are responsible for the appearance of negative momenta and backflow. Then,  sufficient conditions for avoiding these classical effects are $k_{1}\gg 1/a_{x}$ and $R_{0}<f v_{1}/\omega_{x}$. 

\subsection{Quantum backflow}

\begin{figure}
\centerline{\includegraphics[width=0.7\columnwidth]{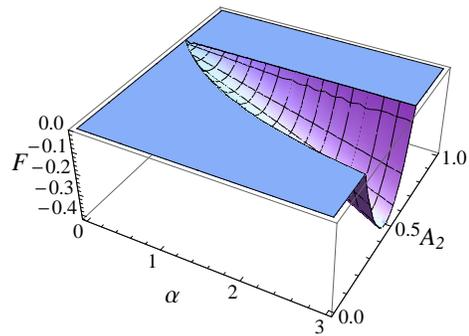}}
\caption{(Color online) Plot of the function $F(\alpha,A_{2})$ defined in Eq. (\ref{eq:fmin0}).
For a given value of $\alpha$, the maximal backflow is obtained for the value $A_{2}$ that minimizes $F$.}
\label{fig:function}
\end{figure}

Let us now turn to the quantum backflow. In order to discuss the optimal setup for having backflow, it is convenient 
to consider the following expression for the current density, 
\begin{eqnarray}
\label{eq:current2}
J_{\Psi}(x,t) &=& \frac{\hbar}{m}|\phi|^{2}\left[q\left(A_{2}^{2}+ A_{1}A_{2}\cos\left(q x + \varphi\right)\right) \right.
\\
&&+ \left.\nabla\theta\left(A_{1}^{2} + A_{2}^{2}+ 2A_{1}A_{2}\cos\left(q x + \varphi\right)\right)\right], 
\nonumber
\end{eqnarray}
that follows directly from Eq. (\ref{eq:current}) and the expression for the total density.
Let us focus on  its behavior around  the center of wave packet, namely at $x\approx(\hbar k_{1}/m)t=d\omega_{x}t$ (the following analysis extends to the whole packet if $d\omega_{x}t$ is much larger than the condensate width). In the asymptotic limit, the phase gradient at the center is $\nabla\theta|_{c}\approx k_{1}$, and the flux in Eq. (\ref{eq:current2}) turns out to be proportional to that of the superposition of two plane waves of 
momenta $k_{1}$ and $k_{2}=k_{1}+q$. This limit is particularly useful because for two plane waves the probability density is a sinusoidal function and the critical density becomes a constant, 
which in practice makes irrelevant the value of the arbitrary phase $\varphi$ we cannot control.
Then, the condition for having backflow at the wave packet center is
\begin{equation}
k_{1}A_{1}^{2} + k_{2}A_{2}^{2}+ (k_{1}+k_{2})A_{1}A_{2}\cos\left(q x + \varphi\right)<0.
\end{equation}
Since all the parameters $k_{i}$ and $A_{i}$ are positive, the minimal condition for having a negative flux is $k_{1}A_{1}^{2} + k_{2}A_{2}^{2} < (k_{1}+k_{2})A_{1}A_{2}$ (for $\cos(\cdot)=-1$),
that can be written as 
\begin{equation}
F(\alpha,A_{2})\equiv 1 + \alpha A_{2}^{2} - (2 + \alpha)A_{2}\sqrt{1-A_{2}^{2}}<0,
\label{eq:fmin0}
\end{equation}
where we have defined $\alpha=q/k_{1}$. The behavior of the function $F(\alpha,A_{2})$ in the region where $F<0$ is depicted in Fig. \ref{fig:function}. 
In particular, for a given value of the relative momentum kick $\alpha$, the minimal value of $F$ is obtained for $A_{2}$ that solves $\partial F/\partial A_{2}|_{\alpha}=0$, that is for
\begin{equation}
2\alpha A_{2}\sqrt{1-A_{2}^{2}} + (2 + \alpha)(2 A_{2}^{2} -1)=0.
\label{eq:minF}
\end{equation}
In order to maximize the effect of backflow and its detection, one has to satisfy a number of constraints. In principle, Fig. \ref{fig:function} shows that the larger the value of  $\alpha=q/k_{1}$, the larger the effect of backflow is. However, $q$ cannot be arbitrarily large as it fixes the wavelength $\lambda=2\pi/q$ of the density modulations, which must be above the current experimental spatial resolution $\sigma_{r}$, $\lambda\gg\sigma_{r}$, for allowing a clean experimental detection of backflow (see later on).  In addition, as discussed before, $k_{1}$ should be sufficiently large for considering negligible the negative momentum components of the initial wave packet, $k_{1}\gg 1/a_{x}$. Therefore, since the maximal momentum that the condensate may acquire after a shift $d$ of the trap is $\hbar k_{1}=m\omega_{x}d$, the latter condition reads $d\gg a_{x}$. By combining the two conditions above, we get the hierarchy $1\ll d/a_{x}\ll(2\pi/\alpha)(a_{x}/\sigma_{r})$. Furthermore,
  we recall that in the interacting case we must have $R_{0}<f v_{1}/\omega_{x}=f d$ in order to avoid classical effects. Therefore, given the value of the current imaging resolution, the non interacting case ($R_{0}\approx a_{x}$) appears more favorable than the TF one (where typically $R_{TF}\gg a_{x}$). Nevertheless, the latter condition can be substantially softened if the measurement is performed at the wave packet center,  away from the left tail where classical effects take place.
\begin{figure}
\centerline{
\includegraphics[width=0.7\columnwidth]{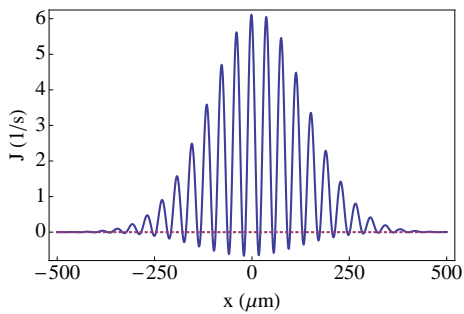}
}
\caption{(Color online) Plot of the flux $J_{\Psi}(x)$.  
Backflow corresponds to $J_{\Psi}<0$. Positions are measured with respect to the wave packet center.}
\label{fig:flux}
\end{figure}

\subsection{An example with $^{7}$Li}

\begin{figure}
\centerline{
\includegraphics[width=0.7\columnwidth]{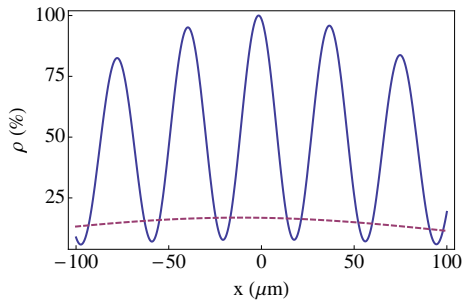}
}
\caption{(Color online) Density (solid) and critical density (dashed) for the case discussed in the text. Backflow occurs in the regions where the density is below the critical value, see Fig. \ref{fig:flux}. Positions are measured with respect to the wave packet center.}
\label{fig:dens}
\end{figure}

As a specific example, here we consider the case of an almost noninteracting $^{7}$Li condensate \cite{salomon} prepared in the ground state of a trap with frequency $\omega_{x}=2\pi\times1~$Hz (yielding  $a_{x}\simeq 38~\mu$m). Then, we shift the trap by $d=80~\mu$m, so that after a time $t_{1}=\pi/(2\omega_{x})=250$ ms the condensate has reached its maximal velocity $\hbar k_{1}/m=\omega_{x}d\simeq 0.5$ mm/s. At this point the axial trap is switched off, and the condensate is let expand for a time $t\gg a_{x}/\omega_{x}d$ until it enters the asymptotic plane wave regime (here we use $t=1$~s).
Finally we apply a Bragg pulse of momentum $\hbar q=\alpha \hbar k_{1}$, with $\alpha=3$, that transfers $24\%$ of the population to the state of momentum $\hbar k_{2}$, according to Eq. (\ref{eq:minF}) ($A_{2}=0.49$, $A_{1}\simeq0.87$) \cite{bragg}. The resulting flux is shown in Fig. \ref{fig:flux}, where the backflow is evident, and  more pronounced at the wave packet center. The corresponding density is displayed in Fig. \ref{fig:dens}, where it is compared with the critical value of Eq. (\ref{eq:denscrit}). The values obtained around the center are $\rho_{\Psi}^{min}\simeq 8\%$ and $\rho_{\Psi}^{crit}\simeq 17\%$.

\subsection{Effects of imaging resolution}

Let us now discuss more thoroughly the implication of a finite imaging resolution $\sigma_r$.
First we note that experimentally it is difficult to obtain a precise measurement of the absolute density, because of uncertainties in the calibration of the imaging setup. Instead, measurements in which the densities at two different points are compared are free from calibration errors and therefore are more precise. Owing to this, it is useful to normalize the total density $\rho_{\Psi}(x,t)=|\phi(x,t)|^{2}\left(A_{1}^{2} + A_{2}^{2}+ 2A_{1}A_{2}\cos\left(qx + \varphi\right)\right)$ to its maximal value $\rho_{\Psi}^{max}\simeq|\phi^{max}|^{2}\left(A_{1}^{2} + A_{2}^{2}+ 2A_{1}A_{2}\right)$. In addition, we have to take into account that, due to the finite resolution $\sigma_{r}$ \cite{resolution}, the  sinusoidal term $\cos\left(qx + \varphi\right)$ is reduced by a factor $\zeta=\exp[-q^{2}\sigma_{r}^{2}/2]$ after the imaging. Then, by indicating with $x_{min}(t)$ the position of the density minima, we have
$$
\left.\frac{\rho_{\Psi}(x_{min},t)}{\rho_{\Psi}^{max}}\right|_{exp}=\frac{|\phi(x_{min},t)|^{2}}{|\phi^{max}|^{2}}
\frac{A_{1}^{2}+ A_{2}^{2} -2\zeta A_{1}A_{2}}{A_{1}^{2}+ A_{2}^{2}+2\zeta A_{1}A_{2}},
$$
where ``exp'' refers to the experimental conditions. 
Instead, the normalized critical density is (close the wave packet center, where $\nabla\theta\approx k_{1}$)
\begin{equation}
\frac{\rho_{\Psi}^{crit}(x,t)}{\rho_{\Psi}^{max}}=\frac{q}{q +2k_{1}}\frac{|\phi(x,t)|^{2}}{|\phi^{max}|^{2}}
\frac{A_{1}^{2} -A_{2}^{2}}{\left(A_{1} + A_{2}\right)^{2}},
\end{equation}
so that, assuming that $|\phi(x,t)|^{2}$ varies on a scale much larger than $\sigma_{r}$ to be unaffected by the finite imaging resolution, the condition for \textit{observing} a density drop below the critical value reads
\begin{equation}
\frac{A_{1}^{2}+ A_{2}^{2} -2\zeta A_{1}A_{2}}{A_{1}^{2}+ A_{2}^{2}+2\zeta A_{1}A_{2}}
=\frac{\alpha}{\alpha +2}
\frac{A_{1} -A_{2}}{A_{1} + A_{2}}.
\end{equation}
In particular, in the example case we have discussed, backflow could be clearly detected with an imaging resolution of about $3~\mu$m, which is within reach of current experimental setups.

\section{conclusions and outlooks}

In conclusion, we have presented a feasible experimental scheme that could lead to the first observation of quantum backflow, namely the presence of a negative flux for states with negligible negative momentum components. By using current technologies for ultracold atoms, we have discussed how to imprint backflow on a Bose-Einstein condensate and how to detect it by a usual density measurement. Remarkably, the presence of backflow is signalled by the density dropping below a critical threshold. Other possible detection schemes could for example make use of \textit{local} velocity-selective internal state transitions in order to spatially separate atoms travelling in opposite directions. Finally, we remark that a  comprehensive understanding of backflow is not only important for its fundamental relation with the meaning of quantum velocity, but as also because of its implications on the use of arrival times as information carriers \cite{muga,Leav}. It may as well lead to interesting applications such as the development of a matter wave version of an optical tractor beam \cite{tractor}, namely, particles sent from a source that could attract towards the source region other distant particles, in some times and locations.

\acknowledgments
M. M. is grateful to L. Fallani and C. Fort for useful discussions and valuable suggestions.
We acknowledge funding by Grants  FIS2012-36673-C03-01, No. IT472-10 and the UPV/EHU program UFI 11/55.  
M. P. and E. T.  acknowledge fellowships by UPV/EHU.


\begin{thebibliography}{10}
\bibitem{note}
This factorization hypothesis allows for a full analytic treatment. Nevertheless, we expect backflow to survive even in case of a generic elongated trap.
\bibitem{bragg2} The duration of the Bragg pulse can be of the order of few hundreds of $\mu$s, therefore very short with respect to timescale of the condensate dynamics.
\bibitem{resolution}
In general, the point-spread-function of the imaging system can be safely approximated with a gaussian function of width $\sigma_r$.
\end{thebibliography}
\end{document}